\documentstyle[12pt,psfig]{article}

\begin{document}

\title{Bounds on Information and the Security of Quantum Cryptography}

\author{Eli Biham$^{(1)}$ and Tal Mor$^{(2)}$}

\date{\today}

\maketitle
(1) Computer Science Department, Technion, Haifa 32000,
Israel; (2) Physics Department, Technion, Haifa 32000,
Israel;

\begin{abstract}

Strong attacks against quantum key distribution 
use quantum memories and 
quantum gates to attack directly the final key.
In this paper we extend a novel security result recently obtained, 
to demonstrate
proofs of security against a wide class of such attacks.
To reach this goal we calculate information-dependent reduced density matrices, 
we study the geometry of quantum mixed state, and we find bounds on the 
information leaked to an eavesdropper.
Our result suggests that quantum cryptography is ultimately secure.

\end{abstract}


Quantum cryptography (e.g.~\cite{BB84,B92}) suggests an  
{\em information secure} 
key distribution. 
It is based on the fact that non-orthogonal quantum 
states cannot be cloned,
and any attempt to obtain information regarding these states
necessarily disturbs them and induces noise.
In principle, the legitimate users of
a quantum key distribution scheme, Alice and Bob,
should quit the protocol
if they notice a noise. 
However, in real protocols, the channels and devices are not perfect,
and some errors are inevitable. 
As long as the rate of errors is small,
these errors must be accepted and corrected by the legitimate
users. As a result, the eavesdropper, Eve, 
can obtain some information on the transmitted data, as long as she
induces less errors than allowed (e.g., by eavesdropping on a small portion
of the transmitted particles).
Furthermore, she can obtain more information using 
the error-correction data transmitted via 
a classical channel.

To overcome these problems, privacy amplification techniques~\cite{BBCM}
were suggested. 
The simplest technique 
uses the parity bit of a long string as the secret bit (where the parity is
zero if the string contains an even number of $1$'s and else it is one).
Such techniques aim to reduce Eve's information on the final key to be
exponentially small with the length of the initial string
(or at least to be much smaller than a single bit).
Unfortunately, a proof of security must stand against an adversary equipped
with {\em any} technology allowed
by the rules of quantum mechanics, and 
neither of the suggested schemes is proven secure (for a different opinion
see~\cite{DM1});
their security 
against sophisticated {\em joint} attacks, which 
use quantum memories, quantum gates, and delayed measurements
to attack {\em directly} the final key,
is only partially established~\cite{Yao,oxford,BM}.
In this work we extend the results of~\cite{BM} much further.

The first hints that privacy amplification might
still be effective against
such attacks were provided by Bennett, Mor and Smolin (BMS)~\cite{BMS}.
Suppose Eve obtains a binary string of $n$ bits where each bit is presented by
non-orthogonal polarization 
states, $\psi_0 = {\cos\alpha \choose \sin\alpha}$ 
or $\psi_1={\cos\alpha \choose -\sin\alpha}$, 
with {\em small} angle $2 \alpha$ between them.
If each bit is measured separately, the optimal 
information on the parity bit
of the string, 
$ I_S(n,\alpha) = (2 \alpha)^{2n}/(2 \ln 2)$,
is exponentially small with $n$.
By measuring all particles together Eve can gain much more information
on the parity bit.
However, the optimal information,
$ I_M(n,\alpha) = c {2k \choose k} \alpha^{2k} $
(with $n=2k$ and $c=1$ for even $n$, and $n=2k-1$ and $c=1/\ln2$
for odd $n$),
is still  
exponentially small with the length of the string.
This result (henceforth, the BMS result) suggests that 
quantum cryptography is secure even when Eve knows
the specification of the privacy amplification technique,
since all privacy amplification techniques are based on similar principles.

In real protocols, Eve does not obtain one of two states with small angle
between them,
but she can probe the states sent from Alice
to Bob using any technique she likes.
Thus, the BMS result only provides some intuition regarding the effectiveness
of privacy amplification.
To make this intuition more adequate to realistic
quantum key distribution protocols,
Biham and Mor~\cite{BM} presented 
a restricted class of joint attacks, called {\em collective attacks},
which can use the BMS method and result:
(a) Eve attaches a  
\underline{separate, uncorrelated} probe to each transmitted
particle using a translucent attack;
(b) Eve keeps the probes in a quantum memory till receiving all classical 
data including error-correction code and privacy amplification data;
(c) Eve performs the optimal measurement on her probes in order to learn
the optimal information on the final key.
The underlined constraints on the probes distinguish the collective 
attacks from more general joint attacks,
and enable analyzing the attacks 
in terms of the density matrices which Eve obtains. 

We concentrate on {\em symmetric collective attacks} 
in which
the same translucent attack is applied to each transmitted
particle, and 
the attack is symmetric to any of the allowed quantum states 
of each particle.
Such an attack induces the same probability of error to each transmitted
bit.
It must be weak, or else it would induce 
a non acceptable error-rate.
Thus, the possible states of Eve's probe cannot differ much.

An explicit example of such a symmetric 
collective attack was presented in~\cite{BM}, together with a proof
of security against it.
In this example Alice and Bob 
use the two state scheme of
Bennett~\cite{B92} (with pure polarization states with angle 
$2 \theta$ between them).
Eve uses, in the first step 
of the collective attack,
the (weak) translucent attack without entanglement~\cite{EHPP} 
(which we call here the EHPP attack),
that leaves each probe in one of two pure states, $\psi_0$ or $\psi_1$, 
with small angle $2 \alpha$ between them.
After an error-estimation step, Alice and Bob 
have an $n$-bit string.
Alice and Bob choose the parity bit of that (full $n$-bit) 
string to be their secret bit,
and Alice sends to Bob some parities of substrings as the error-correction
data\cite{ecc}. 
In~\cite{BM} we calculated 
Eve's density matrices for the parity bit
while taking into account the error-correction data she has~\cite{Dominic}.  
Then, we found Eve's best strategy for measuring the probes
and her optimal mutual information on the parity bit.
For large strings  
and small error-rate (thus, small angle $\alpha$) 
this information decreases exponentially with the length of the string $n$;
e.g., for Hamming codes it is 
\begin{equation}  I(n,\alpha)\le
C(n) (2 \alpha)^{(n+1)/2} \ , \label{BMres} \end{equation}
with $C(n) = \frac{2}{\ln 2 \sqrt \pi}\sqrt{(n+1)}$. 
For a given error-rate, $p_e$, the resultant
angle in the EHPP attack~\cite{therefore}
is $\alpha = (\tan^2(2\theta)\  p_e)^{1/4}\ $, so that
the information $I(n,p_e)$ 
is of the order of $p_e^{(n+1)/8}$. 

In terms of quantum information theory this result (henceforth, the 
BM result) extends the BMS result to the case where parities of
substrings are given (error-correction code).
For purposes of quantum key distribution, the 
BM result provides the first security proof
against a strong attack.
However, it is restricted to attacks in which Eve's probes are in a pure state.
Unfortunately, most possible translucent 
attacks on the two state scheme~\cite{B92}, 
which can be used in the first step
of the collective attack, leave each of Eve's probes in a mixed state.
Also, any translucent attack on the four state 
scheme~\cite{BB84} leaves each probe
in a mixed state (at least for two out of the four possible states).

The aim of this work is to apply the BM result to the  
case of mixed states.
We first demonstrate that any type of information
which can be extracted from certain two-dimensional mixed states can be bounded,
if the solution for pure states is known.
Then we explicitly demonstrate, via two examples, 
how to bound Eve's optimal information
(for a given induced error-rate).
We also calculate the
(individual bit) {\em information-dependent}
reduced density matrices which are in Eve's hands. 

Any state (density matrix) in 2-dimensional Hilbert space
can be written as  
$ \rho = \frac {\hat I + r \cdot \hat\sigma} {2}$ so that
$\rho = \frac{1}{2} \left(\begin{array}{cc} 1 + z & x - i y \\ x + i y & 1 - z 
  \end{array}\right)$,
with $r = (x,y,z)$ being a vector in ${\cal{R}}^3$, $\hat{\sigma}
= (\hat{\sigma}_x, \hat{\sigma}_y, \hat{\sigma}_z)$ the Pauli matrices,
and $\hat I$ the unit matrix. In this ``spin'' notations, each 
state is represented by the corresponding vector $r$.
For pure states $|r| = 1$, and for mixed states $|r|<1$.  
Suppose that $\chi$ and $\zeta$ are two density matrices, 
represented by $r_{\chi}$ and $r_{\zeta}$ respectively.
It is possible to construct the density matrix 
$ \rho = m \zeta + (1-m) \chi \ $
from the two matrices (where $0 \le m \le 1$), and  
the geometric representation of such a density matrix 
$ \rho = \frac{\hat I + r_{\rho} \cdot \hat{\sigma}}{2} \ $ 
is $r_{\rho} = m r_{\zeta} + (1-m) r_{\chi}$. 
Two pure states can always be expressed as
$ | \Phi_0 \rangle = {c \choose s}   $ and  
$ | \Phi_1 \rangle = {c \choose -s}$, with $c=\cos \alpha$ and $s=\sin \alpha$.
Using the notations of density matrices 
$(\Phi_0 \equiv | \Phi_0 \rangle \langle \Phi_0 |$ etc.) the two pure states
are
$ \Phi_p   
= \frac{1}{2} \left(\begin{array}{cc} 1 + z & x \\ x & 1 - z 
  \end{array}\right) $, 
with 
$ z = \cos 2 \alpha $ and $x =  \sin 2 \alpha$ for $p=0$ and
$x = -\sin 2 \alpha$ for $p=1$.
If $\Phi_p$ is used to describe a bit $p$,
the receiver can identify the bit by distinguishing the two pure states.
Two (not necessarily pure) density matrices $\rho_p$ 
in two-dimensional Hilbert space,  
with equal determinants (which are equal to $|r|$) can also be expressed
using similar form with 
$z = |r|\cos 2 \alpha$ and $x = \pm |r| \sin 2 \alpha$.
For two such mixed states 
let us choose a state $\chi_n$, and two 
pure states $\Phi_0$, $\Phi_1$ such that
\begin{eqnarray} \rho_0 &=& m \Phi_0 + (1-m) \chi_n    \nonumber \\   
 \rho_1 &=& m \Phi_1 + (1-m) \chi_n   \label{neutral} \ .  \end{eqnarray}
Let $I$ be some (positive)
measure for the optimal distinguishability of two states,
so that {\em any operation done on them} cannot lead to a 
distinguishability better than $I$.
From the construction of~(\ref{neutral}), 
it is clear that any measure for optimal distinguishability
must find that the two mixed states $\rho_p$ are not more distinguishable than
the two pure states $\Phi_p$ [that is $I(\Phi_0;\Phi_1)\ge I(\rho_0;\rho_1)$]:
Suppose the contrary $I(\Phi_0;\Phi_1)<I(\rho_0;\rho_1)$.
Then, when one receives $\Phi_p$
he can mix them with $\chi_n$ and derive a better distinguishability than
$I(\Phi_0;\Phi_1)$, in contradiction to the definition of 
$I(\Phi_0;\Phi_1)$.

We can choose any measure of an optimal information
carried by these systems 
to describe the distinguishability,
and it should satisfy $I_{\em mixed} \le I_{\em pure}$.
Very complicated types of information can be extracted from such systems,
as for example, the optimal 
information on the parity of an $n$-bit string of 
such quantum bits~\cite{BMS,BM}. 
In the case~\cite{BM}, where parities of substrings 
are given, a solution exists only for pure state with small angles
(the BM result),
and we can now use this known solution 
to bound the optimal information which can be 
extracted from mixed states which are close to each other.
Let $\rho_{\rm cms}$ be the completely mixed state 
$\rho_{\rm cms} = \frac{1}{2} \hat I $.
Also let $\rho_{\downarrow}$ be the pure state of 
spin down in the $z$ direction.
Two cases of eq.~(\ref{neutral}) are useful for our purpose:
(a) $\rho_p = m \Phi_p + (1-m) \rho_{\rm cms} $, 
where the pure states $\Phi_p$ have the same angle as $\rho_p$ (see fig. 1a);
(b) $\rho_p = m \Phi_p + (1-m) \rho_\downarrow $, 
where $\Phi_p$ (which are uniquely determined) are shown in Fig.1b.
The first type of bound is useful if $\rho_p$ have a small angle $\alpha$
(which satisfies $\tan 2 \alpha = x/z$), so that the angle between the pure 
states $\beta = \alpha$ is also small.
The second type of bound is useful when the `distance' $2 x$ between 
the two possible mixed states is small (while $\alpha$ might be large).
In this case $x$ is small and $z$ positive hence the resultant angle 
$\beta$ between the two pure states is small
(following $\tan \beta = \tan 2 \delta = \frac{x}{z+1} \le x$).
Thus, in both cases the angle between the two pure states 
is small so that 
$I(n,\beta)$, eq.~(\ref{BMres}) with an angle $\beta$,
provides an upper bound on 
Eve's information on the final key. 

An explicit calculation of Eve's density matrix as a function 
of $p_e$ must be done separately 
for any suggested attack to obtain $I(n,p_e)$.
However, the fact that Eve is allowed to induce only small error-rate
restricts her possible transformations at the first stage of
the collective attack, hence, the two possible states of each of her probes
must be largely overlapping (for a symmetric attack).
Concentrating on two-dimensional probes, this promises us that the second 
of the two cases above can {\em always}
be used
to bound Eve's information to be exponentially small.
For certain examples -- the first case is sufficient, hence the 
angle between the two possible pure states can be calculated from Eve's
density matrix directly using $\beta = \alpha$.
Let us show two examples in details, to conclude that Eve's information
is exponentially small with the length of the string.
Both examples use the same unitary transformation
but are applied onto different quantum cryptographic schemes,
the two state scheme~\cite{B92} and the
four state scheme~\cite{BB84}. 

In our examples Eve uses a 2-dimensional probe in an initial state
${ 1 \choose 0}$. She performs a unitary transformation
$U {1 \choose 0} |\phi\rangle$ (with $|\phi\rangle$ Alice's state), where 
\begin{equation}    
U = \left(\begin{array}{cccc} 1    &    0     &     0     &    0    \\
                              0    & c_\gamma & -s_\gamma &    0    \\ 
                              0    & s_\gamma &  c_\gamma &    0    \\ 
                              0    &    0     &     0     &    1   
  \end{array}\right) \ , 
\label{wse} \end{equation}
with $c_\gamma = \cos \gamma$, etc.
She chooses a small angle $\gamma$ so that the attack is weak. 
Let Alice's possible initial states be 
$|\phi_p\rangle={\cos \theta \choose \pm \sin \theta}$
in the two state scheme, and 
$|\phi_m\rangle = 
\frac{1}{\sqrt 2}{1 \choose i^m}$ (with $m=0\cdots 3$) in the four state 
scheme. 
The corresponding final states are 
\begin{equation}
|\Psi_p\rangle = \left(\begin{array}{c} \cos\theta     \\
                       \pm    \sin\theta  c_\gamma        \\ 
                       \pm    \sin\theta  s_\gamma        \\ 
                              0    
  \end{array}\right) \ ; \quad
|\Psi_m\rangle = \frac{1}{\sqrt 2} \left(\begin{array}{c} 1  \\
                                     i^m  c_\gamma        \\ 
                                     i^m  s_\gamma        \\ 
                                          0    
  \end{array}\right) \ , 
\label{fin} \end{equation}
respectively.
Bob's reduced density matrices (rdms)
are calculated from $|\Psi\rangle\langle\Psi|$
by tracing out Eve's particle. 
This operation is usually denoted by 
$\rho_B = {\rm Tr}_{{}_E}\ 
[|\Psi\rangle \langle \Psi|]$, where the full formula is given
by eq. 5.19 in~\cite{Peresbook}
($\rho_{nm}=\sum_{\mu\nu} \rho_{n\nu,m\mu} \delta_{\mu\nu}
=\sum_{\mu} \rho_{n\mu,m\mu}$).
We denote this operation by
$\rho_B = {\rm Tr}_{{}_E}\ 
\left[(|\Psi\rangle \langle \Psi|) \hat{I}\right]$, 
where $\hat{I}$ is two dimensional ($\delta_{\mu\nu}$ in eq. 5.19).
From Bob's matrices we find the error-rate,
that is, the probability $p_e$ that he recognizes a wrong bit value.
Calculating Eve's density matrix is more tricky; 
we need to 
take into account the additional information she obtains from the classical
data, in order to obtain an information-dependent rdms.
This is a trivial task for the four-state scheme but a rather confusing 
task in case of the two-state scheme.

In case of the four state scheme Bob measures his particle in one of the 
basis $x$ (corresponding to $m=0,2$) or $y$ ($m=1,3$).
Suppose that Alice and Bob use the $x$ basis; 
Bob's rdms are
$ \rho_B  = \left(\begin{array}{cc} 
 \frac{1}{2} + \frac{1}{2} (s_\gamma)^2     & \pm \frac{1}{2} c_\gamma    \\ 
 \pm  \frac{1}{2} c_\gamma     &   \frac{1}{2} - \frac{1}{2} (s_\gamma)^2     
  \end{array}\right) $, 
leading to an error-rate $p_e=\sin^2(\gamma/2) $ which is the probability 
that he identifies $|\phi_2\rangle$ when $|\phi_0\rangle$ is sent.
Eve has the same knowledge of the basis, hence her 
information-dependent rdms are 
$ \rho_E  = \left(\begin{array}{cc} 
 \frac{1}{2} + \frac{1}{2} (c_\gamma)^2     & \pm \frac{1}{2} s_\gamma    \\ 
  \pm \frac{1}{2} s_\gamma     &   \frac{1}{2} - \frac{1}{2} (c_\gamma)^2     
  \end{array}\right) $, 
so that $x=s_\gamma$, $z=(c_\gamma)^2$, and the relevant angles are 
$2\beta=2\alpha = (\tan)^{-1}(s_\gamma/{c_\gamma}^2)$ (using the first
type of bounds).
For a small angle $\gamma$ we get
$p_e\approx \gamma^2 / 4  + O(\gamma^4)$, 
$\beta  \approx \gamma / 2 + O(\gamma^3)$, and thus 
$p_e \approx \beta^2 + O(\beta^4)$.  The information
is thus bounded by 
$ I(n,p_e) <  C(n) 
(4 p_e)^{(n+1)/4} $  
to be exponentially small
[using eq.~(\ref{BMres})].   

In case of the two-state scheme 
Bob's rdms are
$  \rho_B  = \left(\begin{array}{cc} 
 (c_\theta)^2 + (s_\theta)^2 (s_\gamma)^2  & 
           \pm   c_\theta s_\theta c_\gamma    \\ 
\pm c_\theta s_\theta c_\gamma    & (s_\theta)^2 (c_\gamma)^2  
  \end{array}\right) $. 
Bob chooses one of two possible measurements, $M_{0\rightarrow 1}$ or
$M_{1\rightarrow 0}$, with equal probability $p_{0\rightarrow 1}
=p_{1\rightarrow 0} = 1/2$;
In case of $M_{0\rightarrow 1}$, Bob measures the received state to distinguish 
$\phi_0$ from its orthogonal state ${\phi_0}'$ 
and finds a conclusive result `1'  
whenever he gets ${\phi_0}'$. (The conclusive result `0' is obtained
by replacing $0$ and $1$ in the above).
The error-rate is the probability of identifying
${\phi_p}'$ when $\phi_p$ is sent, and  it is 
$p_e = (s_\theta)^2 (c_\theta)^2 [ 1 - c_\gamma]^2 + (s_\theta)^4 (s_\gamma)^2$.

To obtain Eve's density matrices one must take into account all 
the information she possibly has.
If one ignores the classical information and calculates the standard 
rdms (as in~\cite{FP}), then 
the result is of significant  
importance to quantum information,
while it 
is less relevant to quantum cryptography.
Recall that Bob keeps only
particles identified conclusively (as 
either ${\phi_0}'$ or ${\phi_1}'$);
Bob informs Alice --- and thus Eve --- which they are, and, as a result,
Eve knows that Bob received either 
${\phi_0}'$ or ${\phi_1}'$ in his measurement, and not
${\phi_0}$ or ${\phi_1}$.
This fact influences her density matrices, 
and these are not given anymore by the simple 
tracing formula  
$\rho_E = {\rm Tr}_{{}_B}\ 
\left[(|\Psi\rangle \langle \Psi|) \hat{I}\right]$. 
In general, {\em information dependent} rdms are obtained  
by replacing $\hat{I}$ by any other positive operator $\hat{A}$:
\begin{equation} \rho_E = {\rm Tr}_{{}_B}\ 
\left[(|\Psi\rangle \langle \Psi|) \hat{A} 
\right] \label{tracing}     \end{equation}
(This is a rather obvious conclusion from the discussions prior to eq.
5.19 and also from 
page 289 in sec. 9 in~\cite{Peresbook};
The correctness of this technique can easily be verified~\cite{Verify}).
In our case 
$ \rho_E = {\rm Tr}_{{}_B}\left[ ( |\Psi \rangle \langle \Psi |)  
(\frac{1}{2} |{\phi_0}'\rangle \langle {\phi_0}'| + 
\frac{1}{2} |{\phi_1}'\rangle \langle {\phi_1}'|) \right]$,
where the halves result from $p_{0\rightarrow 1}$ and 
$p_{1\rightarrow 0}$.
This tracing technique leads to 
\begin{equation}    
\rho_E =  
    \left(\begin{array}{cc} 
 (s_\theta)^2 (c_\theta)^2 + (s_\theta)^2 (c_\theta)^2 (c_\gamma)^2  & 
                                 \pm    c_\theta (s_\theta)^3  s_\gamma    \\ 
 \pm c_\theta (s_\theta)^3  s_\gamma & 
                                        (s_\theta)^4 (s_\gamma)^2  
  \end{array}\right)  . 
\label{2-sE} \end{equation}
After normalization we get
$x = 2 s_\gamma c_\theta (s_\theta)^3 /{\rm Tr} \rho_E$ and 
$z = \frac{1+z}{2} - \frac{1-z}{2} = 
[(c_\theta)^2 (s_\theta)^2 [1+(c_\gamma)^2] - (s_\theta)^4 (s_\gamma)^2) /
{\rm Tr} \rho_E$.
The relevant angles are again $2\beta = 2 \alpha = \tan^{-1} (x/z)$.
For small angle $\gamma$ we get  
$p_e \approx {s_\theta}^4 \gamma^2 + O(\gamma^4) $,
$2 \beta  \approx (s_\theta / c_\theta ) \gamma + O(\gamma^3)$.
Finally we get 
$p_e \approx (s_\theta)^2 (c_\theta)^2 (2 \beta)^2 + O(\beta^4)$ 
from which $I(p_e,n)$ can be easily calculated as in the previous example.

The information available to Eve when she performs any other
symmetric collective
attack with two-dimensional probes can also be calculated using our method.
Although we do not know to find the optimal attack of that type yet,   
our method can still prove security against it, since
there is some (small enough)
error-rate, such that Eve's probes have small angles between them, 
and thus, our proof can be applied. The second type of bounds is usually
irrelevant when the 
attack is given 
(since Eve's initial state is usually in pure state), 
but it can be very useful for finding 
the optimal attack, requiring only to find the maximal `distance', 
$2 x$, between
the two possible states of the probe.

More general collective attacks can use non-symmetric translucent attacks
and/or can use probes in higher dimensions,
in the first step of the collective attack.
Methods similar to ours can be used for proving security against various 
non-symmetric collective attacks (in 2-dimensions), but the calculation
becomes more complicated and is beyond the goals of this work.
Our bounds cannot be used when 
Eve uses higher dimensional probes.
Indeed, in this case the two possible states of
each probe are still highly overlapping, and the same intuition
which holds in our paper shall still hold.  However, 
extending the information bounds we found to three or four dimensions
might be a difficult task (such analysis of
dimensions higher than four is not required since they cannot help
the attacker due to  
the reasons shown in~\cite{FP}).

A more crucial issue is the possibility of finding stronger joint attacks  
which are not collective. 
Let us present the strong argument which is the basis for approaching 
the security problem through the collective attack:
by the time Eve holds the transmitted particles she has no knowledge of the 
error-correction and privacy amplification techniques to be used
by Alice and Bob.
She even doesn't know which particles will be discarded in the 
error-estimation stage, and how the common bits will be reordered.
Thus, we conjecture that 
she cannot gain information by searching or by creating correlations between 
the transmitted particles; she better keep one separate probe for each 
particle, and perform the measurements after obtaining the missing 
information as is done in the collective attacks. 
Any attempt of creating such coherent correlations at the first
step of the attack induces error, while it cannot lead 
to an increase in the resultant information;
indeed it could help Eve if she would guess correctly the
required correlations (e.g., the final string, from which the parity is 
calculated), but the probability of successful guess
is exponentially small.
Unfortunately, proving this intuitive argument is yet an open problem.
 
It is a pleasure for us to thank Asher Peres, William Wootters
and Gilles Brassard
for helpful discussions.

\newpage
\begin{figure}[p]
\hspace{1cm}
\psfig{file=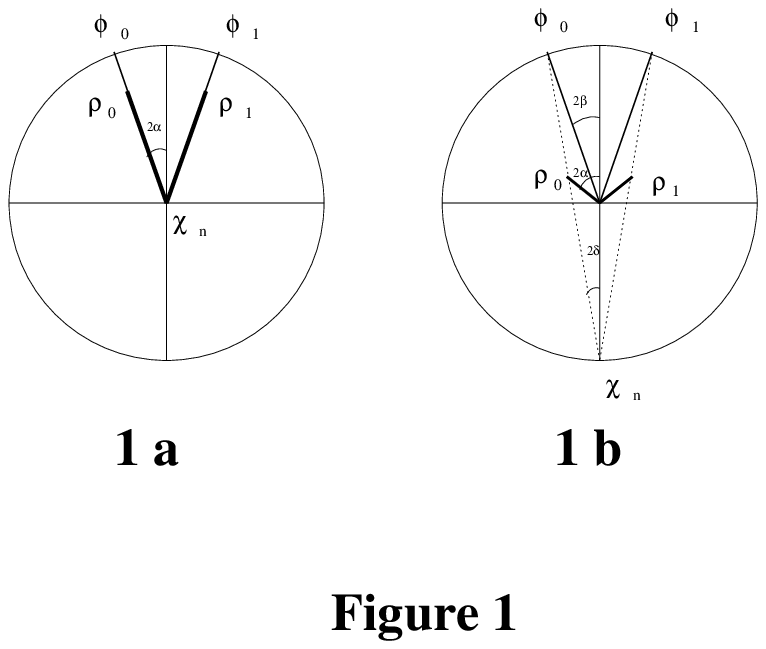}
\caption{Two ways of constructing the two density matrices $\rho_p$ from
two pure states $\Phi_p$ and a third state $\chi_n$ 
common to both density matrices.
In a), $\chi_n = \rho_{cms}$, the completely mixed state.
In b), $\chi_n = \downarrow_z$, the ``down $z$'' pure spin state.}
\end{figure}

\begin{thebibliography}{99}
\bibitem{therefore}
However, it is easy to check that $\alpha$ is of order ${p_e}^{1/2}$
if the EHPP attack is applied on various other states  
which are more sensitive to this specific eavesdropping attack.
We emphasize this point to avoid the impression that 
the translucent attack we use later on is inferior to the EHPP attack.
\bibitem{Verify} For instance, let a singlet state be shared by Eve  
and Bob, and let Bob find his particle in a state 
$|\phi\rangle ={\cos \eta \choose \sin \eta}$.
Replacing $\hat I$ by $|\phi\rangle \langle\phi|$ as in~(\ref{tracing}),
we see that the
state in Eve's hands is not $\rho_{cms}$ but
${\sin\eta \choose -\cos\eta}$ as expected.
\end{thebibliography}
\end{document}